\documentclass[twocolumn]{aastex63}%
\usepackage{lineno}
\usepackage{amsmath}
\usepackage{mathrsfs}
\usepackage{verbatim}
\usepackage{longtable}
\usepackage{cuted}

\submitjournal{ApJ}
\shorttitle{GRB~221009A in the Wind Environment}
\shortauthors{Ren et al.}
\begin{document}
\title{The Possibility of Modeling the Very High Energy Afterglow of GRB~221009A in a Wind Environment}

\correspondingauthor{Jia Ren}
\email{jia@smail.nju.edu.cn}
\correspondingauthor{Zi-Gao Dai}
\email{daizg@ustc.edu.cn}

\author[0000-0002-9037-8642]{Jia Ren}
\affiliation{School of Astronomy and Space Science, Nanjing University, Nanjing 210023, China}
\affiliation{Key Laboratory of Modern Astronomy and Astrophysics (Nanjing University), Ministry of Education, Nanjing 210023, China}
\author[0000-0002-8385-7848]{Yun Wang}
\affiliation{Key Laboratory of Dark Matter and Space Astronomy, Purple Mountain Observatory, Chinese Academy of Sciences, Nanjing 210034, China}
\affiliation{Department of Astronomy, School of Physical Sciences, University of Science and Technology of China, Hefei 230026, China}
\author[0000-0003-0726-7579]{Lu-Lu Zhang}
\affil{Guangxi Key Laboratory for Relativistic Astrophysics, School of Physical Science and Technology, Guangxi University, Nanning 530004, China}
\author[0000-0002-7835-8585]{Zi-Gao Dai}
\affiliation{Department of Astronomy, School of Physical Sciences, University of Science and Technology of China, Hefei 230026, China}
\affiliation{School of Astronomy and Space Science, Nanjing University, Nanjing 210023, China}

\begin{abstract}
In this paper, we model the dynamics and radiation physics of the rarity event GRB~221009A afterglow in detail.
By introducing a top-hat jet that propagates in an environment dominated by stellar winds,
we explain the publicly available observations of afterglow associated with GRB~221009A over the first week.
It is predicted that GRB~221009A emits a luminous very high energy (VHE) afterglow
based on the synchrotron self-Compton (SSC) process in our model.
We show the broadband spectral energy distribution (SED) analysis results of GRB~221009A,
and find that the SSC radiation component of GRB 221009A is very bright in the $0.1-10$~TeV band.
The integrated SED shows that the SSC emission in the TeV band
has significantly higher than the detection sensitivity of LHASSO, MAGIC and CTA.
However, since the release of further observations,
deviations from the standard wind environment model gradually show up in data.
For example, the late-time multiband afterglow cannot be consistently explained
under the standard wind environment scenario.
It may be necessary to consider modeling with a structured jet
with complex geometry or a partial revision of the standard model.
Furthermore, we find that the inclusion of GeV observations could break the degeneracy between model parameters,
highlighting the significance of high-energy observations in determining accurate parameters for GRB afterglows.
\end{abstract}

\keywords
{Gamma-ray bursts (629); High energy astrophysics (739)}
\section{Introduction}
On 2022~October~9 at 13:16:59~UT, the {\em Fermi} Gamma-Ray
Burst Monitor (GBM) triggered long duration gamma-ray burst GRB~221009A \citep{GCN32636}.
Not too late, at 14:10:17~UT,
The {\em Swift} Burst Alert Telescope (BAT) triggered GRB~221009A, aka Swift J1913.1+1946 \citep{GCN32632}.
The {\em Fermi} Large Area Telescope (LAT) detected high-energy emission from GRB~221009A at 14:17:05.99 \citep{GCN32637},
and the high energy photon observed by {\em Fermi}/LAT
reaches 99.3~GeV \citep{GCN32658}, even 397.7~GeV \citep{GCN32748}.
The {\em Swift} X-ray Telescope (XRT) have fruitful follow up observations \citep{GCN32731}.
GRB~221009A is extraordinary bright with isotropic prompt emission energy
$E_{\gamma,\rm iso}\simeq1.5\times 10^{55}$~erg
\citep{An_2023_arXiv230301203A,Yang_2023_Zhao_arXiv230300898Y}
with redshift $z=0.151$ \citep{GCN32648,GCN32686,Malesani_2023_Levan_arXiv230207891M}.
Very energetic GRBs at such close distances are estimated to occur only once in a century \citep{GCN32793}
or even ten thousand years \citep{Burns_2023_Svinkin_arXiveprints_v.p2302..14037arXiv},
and it is possible to detect very high energy (VHE) photons \citep{xue2009very}.
As expected, China's Large High Altitude Air Shower Observatory (LHAASO)
recorded at least tens of thousands of photon signals in the VHE region above 100~GeV,
including photons with energy greater than 10~TeV \citep{GCN32677},
which can give the finest measurement of the lightcurve in the highest energy band of GRBs.

In recent years, several long GRBs have been successfully observed corresponding VHE afterglows.
Long GRBs with the same VHE afterglows as GRB~221009A,
including GRBs~180720B, 190114C, 190829A, 201015A, and 201216C,
have four orders of magnitude on $E_{\gamma,\rm iso}$
from $\sim 10^{50}~\rm erg$ to $\sim 10^{54}~\rm erg$.
The VHE afterglow of GRB~180720B ($E_{\rm\gamma,iso}=6\times 10^{53}$~erg at $z=0.654$)
were firstly detected with the High Energy Stereoscopic System (H.E.S.S.)
in a confidence level of $5.3\sigma$ in the $0.1-0.4$~TeV band
\citep{Abdalla_2019_Adam_Nature_v575.p464..467}.
Moreover, the VHE afterglow of GRB~190114C ($E_{\rm\gamma,iso}=3\times10^{53}$~erg
at $z=0.4245$) was convincingly detected with
the Major Atmospheric Gamma Imaging Cerenkov (MAGIC) telescopes in the $0.3-1$~TeV band
in a high confidence level of $> 50\sigma$ \citep{MAGIC_Collaboration_2019_Acciari_Natur.575..459M}.
The TeV afterglow of nearby GRB 190829A was detected
($E_{\rm\gamma,iso}=2\times 10^{50}$~erg at $z=0.0785$) with H.E.S.S. in the $0.2-4$~TeV band
with a confidence level of $21.7\sigma$
\citep{HESS_Collaboration_2021_Abdalla_Sci...372.1081H}.
The VHE afterglows of GRBs~201015A
($>140$~GeV, $3.5\sigma$, \citealp{Suda-2022-Artero-icrc.confE.797S})
and 201216C ($\sim 100$~GeV, $6\sigma$, \citealp{Fukami-2022-Berti-icrc.confE.788F})
were detected by the MAGIC telescopes also, but the data have not yet been released.
These two bursts have prompt emission energy
$E_{\rm\gamma,iso}=1.1\times10^{50}$~erg with the redshift $z=0.423$ of GRB~201015A,
and $E_{\rm\gamma,iso}=5.76\times10^{53}$~erg with the redshift $z=1.1$ of GRB 201216C,
respectively (Zhang et al. 2022, in preparation).
Since the extragalactic background light (EBL)
absorbs sub-TeV/TeV photons from high-redshift sources,
the VHE afterglows of low-redshift GRBs should be easier to detect
(e.g., \citealp{Finke_2010_Razzaque_apj_v712.p238..249,Dominguez_2011_Primack_mnras_v410.p2556..2578}).
Overall, it seems that both energetic and sub-energetic GRBs
can accelerate the particles to an extremely high energy and produce the VHE afterglows
(e.g. \citealp{Hurley_1994_Dingus_Natur.372..652H, Murase_2006_Ioka_ApJ...651L...5M}).

The new window opened on TeV emission of GRBs usher in a new era of GRB physics at extremely high energy.
The VHE afterglows of GRBs are generally believed to be attributed to synchrotron,
synchrotron self-Compton (SSC), and/or external inverse-Compton (EIC) radiations
of the electrons accelerated in relativistic jets (e.g., \citealp{Dermer_2000_Chiang_ApJ...537..785D,
Zhang_2001_Meszaros_ApJ...559..110Z, Sari_2001_Esin_apj_v548.p787..799,
He_2009_Wang_apj_v706.p1152..1162,MAGIC_Collaboration_2019_Acciari_Natur.575..459M,
Acciari-2021-Ansoldi-ApJ...908...90A,HESS_Collaboration_2021_Abdalla_Sci...372.1081H,
Zhang-2021-Ren-ApJ...917...95Z,Zhang_2021_Murase_apjl_v908.p36..36L}).
Besides, the hadronic processes may also play an important role
\citep[e.g.,][]{Sahu_2022_Valadez_ApJ...929...70S}.
To distinguishing between these components,
accurate modeling of broadband spectral energy distributions (SEDs) is required.
Thus, abundant and high-quality observations from radio to TeV at same timeslices for SEDs
and continuously monitored multiband afterglow lightcurves
will provide a powerful support for understanding the physics of GRB radiation
(e.g., \citealp{Wang_2019_Liu_apj_v884.p117..117,Zhang_2020_Christie_mnras_v496.p974..986,
Joshi_2021_Razzaque_mnras_v505.p1718..1729,Yamasaki_2022_Piran_mnras_v512.p2142..2153}).

It is widely accepted that long GRBs can occur during the core collapse of stars.
Accordingly, stars may have undergone a drastic material-loss process before finally collapsing,
resulting in circumburst environments dominated by stellar winds
\citep{Dai_1998_Lu_mnras_v298.p87..92,Chevalier_2000_Li_apj_v536.p195..212a}.
Previous works have inferred that
GRB~190114C maybe occur in stellar wind environment,
but the other possibilities have not been ruled out
(e.g., \citealp{Asano_2020_Murase_ApJ...905..105A, Joshi_2021_Razzaque_MNRAS.505.1718J}).
In this work, we find that the afterglow lightcurve and SED of GRB~221009A
may support this occurrence.
Although other possibilities cannot be ruled out, such as a structured jet,
our work firstly provides a interpretation of the observations.

This paper is organized as follows.
In Section~{\ref{obsdata}}, we briefly describe the analysis and collection of observational data.
In Section~{\ref{model}}, we introduce our model considerations in detail
and present the results of model parameter inference.
We summarize and discuss the significance of our results in Section~{\ref{Summary}}.
We take the cosmology parameters as $H_0=67.8~\rm km~s^{-1}~Mpc^{-1}$, and $\Omega_M=0.308$ \citep{PlanckCollaboration_2016_Ade_aap_v594.p13..13A}.

\section{Multiwavelength Observation and spectral analysis of GRB 221009A}\label{obsdata}
\subsection{Fermi/{\rm LAT} data analysis}
{\em Fermi}/LAT is a pair conversion telescope covering a wide energy band
(from 20 MeV to greater than 300 GeV, \citealp{atwood2009large}).
We analyze the observation data of {\em Fermi}/LAT in the direction of GRB~221009A
(RA = 288.282, Dec = 19.495, from \cite{GCN32658} based on the standard
procedure\footnote{\url{https://fermi.gsfc.nasa.gov/ssc/data/analysis/scitools/}}.
Considering the duration and high-energy events of GRB~221009A \citep{GCN32748},
data were extracted with the energy range of 100~MeV to 1~TeV,
using transient event class (evclass=16) before 400~seconds
and source event class (evclass=128) after that,
and the corresponding instrument response functions (IRFs) file
are used in the unbinned likelihood analysis.
We perform these analyses by using the {\tt Fermitools} package and use {\tt make4FGLxml} from user contributions\footnote{\url{https://fermi.gsfc.nasa.gov/ssc/data/analysis/user/}} to generate initial model files. For bins with TS values $< 9$, we use the {\tt UpperLimits} tool to calculate the upper limit of flux.
Considering possible pile-up effects \citep{GCN32760} and the caveats given by $Fermi$\footnote{\url{https://fermi.gsfc.nasa.gov/ssc/data/analysis/grb221009a.html}}, we ignore the LAT data of GBM trigger time\citep{GCN32636} $T_0$ - $T_0$ + 294 seconds.

\subsection{Other follow-up observations observations}
We have collected follow-up observations that have been reported from
GCN\footnote{\url{https://gcn.gsfc.nasa.gov}} or
ATel\footnote{\url{https://www.astronomerstelegram.org}}.
For the {\em Swift}/XRT observation data, we rebin the 0.3-10~keV light curve
based on the counts per bin (PC=600, WT=200)\footnote{\url{https://www.swift.ac.uk/xrt\_curves/}},
from about 3300~seconds to 2.3~days after {\em Fermi}/GBM trigger.
We take into account the information of the time-resolved spectrum
into the converting count-rates to energy fluxes from the spectral evolution data
given by the ``burst analyzer'' tool\footnote{\url{https://www.swift.ac.uk/burst\_analyser/}}
(see more details in \citealp{evans2007online,evans2009methods}).
And, the {\em Swift}/XRT spectrum in the different time slices we used
are extracted by an online analysis tool\footnote{\url{https://www.swift.ac.uk/xrt\_spectra/}}.
We have collected the optical data ($r$, $r'$, $R$ and $z$-band, \citealp{
GCN32644, GCN32645, GCN32646, GCN32652, GCN32659, GCN32662,
GCN32667, GCN32669, GCN32670, GCN32678, GCN32684, GCN32692,
GCN32693, GCN32709, GCN32729, GCN32743, GCN32752, GCN32758,
GCN32771, GCN32759, ATel15665})
and radio data (frequencies around 1.5~GHz, 5~GHz, 15~GHz and 97.5~GHz, \citealp{
GCN32653, GCN32655, GCN32676, GCN32700, GCN32740, GCN32761, ATel15671})
for our model analysis.
In addition, the High-Altitude Water Cherenkov Observatory (HAWC)
has reported their observation result started from
$\sim 8$~hours after the trigger time and lasted $\sim 3.4$~hours.
Assuming a power law spectrum with photon index of $-2.0$,
they found no significant detection and proceeded to calculate the 95\% upper limit
on the flux at 1~TeV as $4.16\times 10^{-12}~\rm TeV~cm^2~s^{-1}$ \citep{GCN32683}.

\begin{figure}[htbp]
 \centering
\includegraphics[angle=0,width=0.45\textwidth]{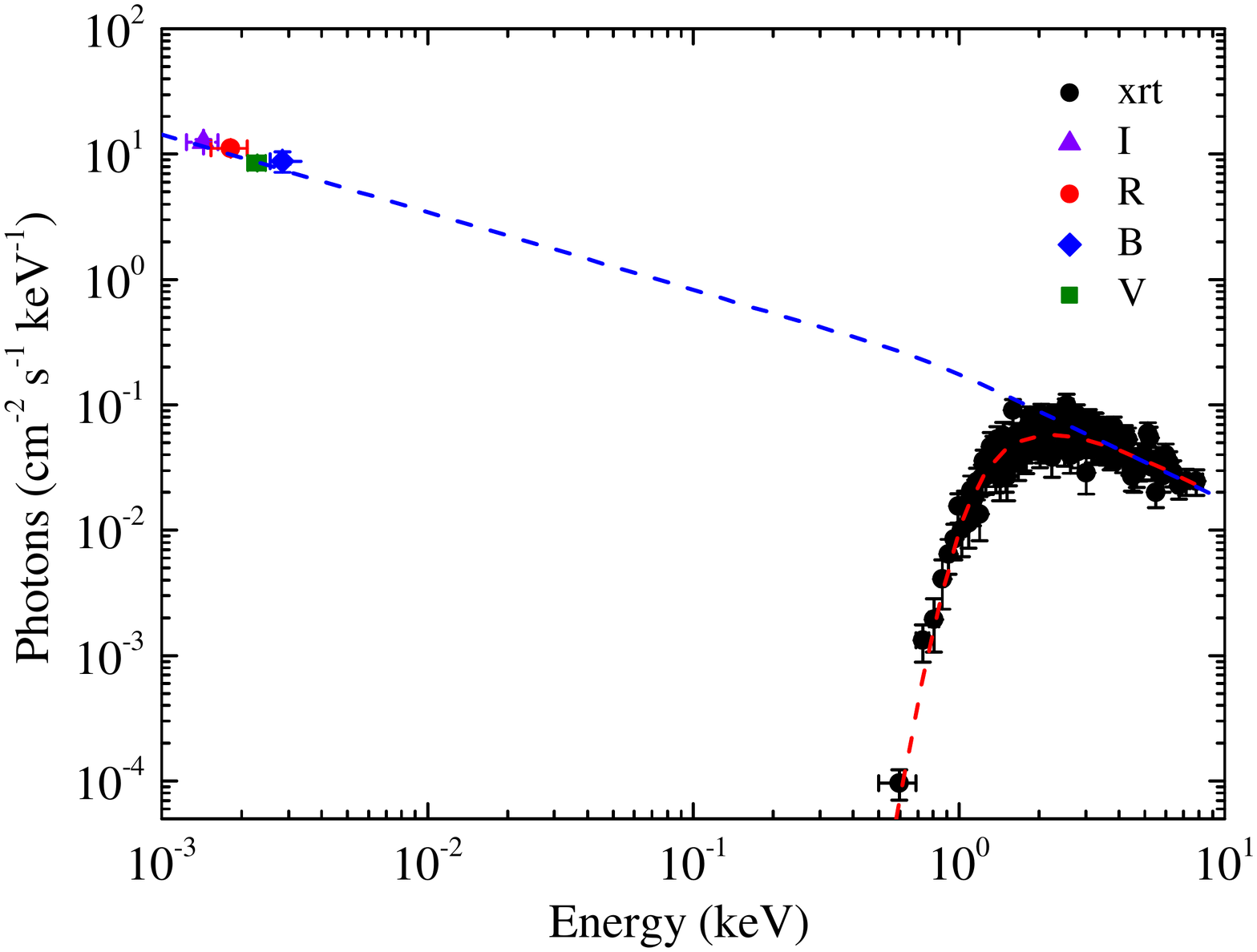}
\caption{Joint optical-X-ray afterglow observed in the time
interval of $[5.3\sim 5.7]\times 10^4$~seconds of GRB~221009A
along with our fit with a broken power-law function (the red dashed line).
The optical data are extinction-corrected for our Galaxy.
The \ion{H}{1} absorption of our Galaxy is also considered in our spectral fit.
The blue dashed line is plotted by using the model parameters of photon indices and break energy.}
\label{joint spectral}
\end{figure}

\subsection{Joint spectral analysis of X-ray and optical band}
We have checked whether or not the host galaxy extinction exists  at the optical band.
The joint spectral analysis of X-ray and optical afterglows at the timeslice
from $5.3\times10^{4}$ to $5.7\times10^{4}$~seconds is using {\tt XSPEC} (version 12.12.1) package.
The above epoch including $R$, $I$, $B$, $V$ bands and they are AB magnitude
except for the $B$ band \citep{GCN32659, GCN32669}.
The Galactic extinctions of four bands are
$A_R=3.339$, $A_I=2.317$, $A_V=4.221$ and $A_B=5.582$, respectively\footnote{
\url{https://ned.ipac.caltech.edu/forms/calculator.html}}.
\citep{Schlafly_2011_Finkbeiner_ApJ...737..103S}.
The Galactic \ion{H}{1} column density
$N_{\rm H}^{\rm Gal}= 5.36\times 10^{21}~{\rm cm}^{-2}$
at the GRB direction is fixed.
The extinction curve of the host galaxy is taken as the same of Mike Way (MW)
\citep{Pei_1992_ApJ...395..130P} by setting $R_V=3.08$.
We find that the extinction value of the host galaxy with the absorption model is very small and negligible.
Finally, we fit the joint spectrum of the X-ray-optical afterglow with a broken power-law model
and the derived photon index are $\Gamma_{\rm O,X}^{1}=1.62\pm0.17$
and $\Gamma_{\rm O,X}^{2}=2.02\pm0.08$.
The break energy is $E_{\rm break}=(0.79\pm 2.18)~\rm keV$.
The reduced $\chi^{2}$ is $141.03/143=0.99$.
The fitting result is shown in Figure~\ref{joint spectral}.
Currently, some works present results of the extinction values of GRB~221009A
and find some significant influence from the host galaxy
\citep[e.g.,][]{Kann_2023_Agayeva_arXiv230206225K,
Levan_2023_Lamb_arXiv230207761L,Williams_2023_Kennea_arXiv230203642W}.
Our results may differ due to the lack of UV band in the data we used,
owing to the broken power-law spectral index differing from the prediction
based on synchrotron radiation mechanism.

\section{physical implication}\label{model}
\subsection{Afterglow model Inference}
We try to explain the multiband observations of GRB~221009A
from radio to GeV band and infer the physical parameters through the afterglow model.
The afterglow model is based on a Python-Fortran hybrid code {\tt ASGARD} package we developed
and can be modified for different physical considerations (e.g., \citealp{Ren_2020_Lin_apj_v901.p26..26L,Zhang-2021-Ren-ApJ...917...95Z}).
In this work, we consider the synchrotron radiation and SSC processes of electrons
in the jet to generate realistic radiation behavior.
We briefly describe the main numerical methods and model used as follows.

The dynamics of the external-forward shock of the jet described as
\citep{Nava_2013_Sironi_mnras_v433.p2107..2121,Zhang_2018pgrb.book.....Z},
\begin{widetext}
\begin{equation}\label{eq:dGamma}
\frac{d\Gamma}{dr}=
-\frac{\Gamma(\Gamma^2-1)(\hat{\gamma}\Gamma-\hat{\gamma}+1)\frac{dm}{dr}c^2
-(\hat{\gamma}-1)\Gamma(\hat{\gamma}\Gamma^2-\hat{\gamma}+1)(3U/r)}
{\Gamma^2[m_0+m]c^2+(\hat{\gamma}^2\Gamma^2-\hat{\gamma}^2+3\hat{\gamma}-2)U}~,
\end{equation}
\end{widetext}
\begin{equation}\label{eq:diff_U}
\frac{dU}{dr}=(1-\epsilon)(\Gamma-1) c^{2}\frac{dm}{dr}
-(\hat{\gamma}-1)\left(\frac{3}{r}-\frac{1}{\Gamma} \frac{d \Gamma}{d r}\right) U~,
\end{equation}
where ${dm}/{dr}=n(r) m_p r^{2}$
with $n(r)$ being the particle density of
circum-burst medium and $m_p$ being the proton mass,
and $\Gamma(r)$, $m(r)$, $U(r)$, and $\epsilon$
are the bulk Lorentz factor, the swept-up mass,
the internal energy, and the radiation efficiency
of electrons in the external-forward shock, respectively.
The adiabatic index is $\hat{\gamma} \simeq
(5-1.21937\zeta+0.18203 \zeta^{2}
-0.96583 \zeta^{3}+2.32513 \zeta^{4}
-2.39332 \zeta^{5}+1.07136 \zeta^{6})/3$
with $\zeta \equiv \Theta /(0.24+\Theta)$,
$\Theta \simeq
(\Gamma \beta/3)
\left[\Gamma \beta+1.07(\Gamma \beta)^{2}\right]/
\left[1+\Gamma \beta+1.07(\Gamma \beta)^{2}\right]$,
and $\beta = \sqrt{1-1/\Gamma^2}$
(\citealp{PeEr_2012__apjl_v752.p8..11}).
We have numerically solved equations~(\ref{eq:dGamma}) and~(\ref{eq:diff_U})
with the forth-order Runge-Kutta method.
$\epsilon_e$ and $\epsilon_B$ are the equipartition factors for the energy in electrons and
magnetic field in the shock, respectively.
Then, the magnetic field behind the shock
is $B^{\prime}={[32\pi { \epsilon_B}{n(r)}]^{1/2}}\Gamma c$,
where ``$\prime$" marks the co-moving frame of shock.
The swept-in electrons by the shock are accelerated
to a power-law distribution of Lorentz factor $\gamma_e$,
i.e., $Q\propto {\gamma'_e}^{ - p}$ for
$\gamma'_{e, \min}
\leqslant \gamma_e \leqslant \gamma'_{e,\max}$,
where $p (>2)$ is the power-law index,
$\gamma_{e,\min}=\epsilon_e(p-2)m_{\rm p}\Gamma/[(p-1)m_{\rm e}]$
(\citealp{Sari_1998_Piran_apj_v497.p17..20L}),
and $\gamma_{e,\max}=\sqrt{9m_e^2{c^4}/[8B'{q_e}^3(1+Y)]}$
with $q_e$ being the electron charge \citep{Kumar_2012_Hernandez_mnras_v427.p40..44L},
where $Y$ is the Compton parameter.
Then, one can have $\epsilon=\epsilon_{\rm rad}\epsilon_e$
with $\epsilon_{\rm rad}=\min \{1,(\gamma_{e,\min}/\gamma_{e,\rm c})^{(p-2)}\}$
(\citealp{Sari_2001_Esin_apj_v548.p787..799,Fan_2008_Piran_mnras_v384.p1483..1501}),
where $\gamma_{e,\rm c}=6 \pi m_e c/[\sigma_{\rm T}\Gamma {B'}^2 t'(1+Y)]$
is the efficient cooling Lorentz factor of electrons with $\sigma_{\rm T}$ being the Thomson scattering cross section.

We denote the instantaneous electron spectrum as ${dN_{e}}/{d\gamma_{e}^{\prime}}$,
of which the evolution can be solved based on the continuity equation of electrons,
\begin{equation}
\frac{\partial}{\partial{t'}}\left(\frac{dN_e}{d\gamma'_e}\right)
+\frac{\partial}{\partial\gamma'_e}\left[\dot{\gamma}'_{e, \rm tot}\left(\frac{dN_e}{d\gamma'_e}\right)\right]
= Q\left(\gamma'_e,t'\right)~,
\end{equation}
We refer to \cite{Fan_2008_Piran_mnras_v384.p1483..1501} for the way to solve this equation.
We note that the Compton parameter $Y(\gamma_{e}^{\prime})$
has been solved based on the work of \cite{Fan_2006_Piran_mnras_v369.p197..206}.
On the numerical method, we solve the continuity equation using a finite difference method of
the third-order total-variation-diminishing Runge-Kutta method of time $t'$ (TVD+RK3)
and fifth-order weighted essentially non-oscillatory method
of $\gamma_e^{\prime}$ (WENO5, \citealp{Jiang_1996_Shu_JCoPh.126..202}).

In the X-ray/optical/radio bands,
the main radiation mechanism of the electrons in GRB jets
is synchrotron radiation
(\citealp{Sari_1998_Piran_apj_v497.p17..20L,Sari-1999-Piran-ApJ...517L.109S}).
The spectral power of synchrotron radiation of $n'_{e}(r,{\gamma'_e})$ at a given frequency $\nu'$ is
\begin{equation}
P'(\nu',r)=\frac{\sqrt{3}q_e^3B'}{m_e c^2}\int_{0}^{\gamma'_{e,\rm max}}
F\left(\frac{\nu'}{\nu'_{\rm c}}\right)n'_e(r,{\gamma'_e})d\gamma'_e~,
\end{equation}
where $F(x)=x\int_{x}^{+\infty}K_{5/3}(k)dk$ with $K_{5/3}(k)$
being the modified Bessel function of 5/3 order
and $\nu'_{\rm c}=3q_e B'{\gamma'_e}^2/(4\pi m_e c)$.

The emission of the SSC process is calculated based on the electron spectrum
and seed photons from the synchrotron radiation
\citep[e.g.,][]{Geng_2018_Huang_apjs_v234.p3..3, Huang-2022-ApJ...931..150H}.
We have numerically solved the Klein-Nishina effect and the $\gamma\gamma$ annihilation effects
(e.g., \citealp{Gould_1967_Schreder_PhRv..155.1404, Fan_2008_Piran_mnras_v384.p1483..1501,Nakar_2009_Ando_apj_v703.p675..691,
Murase_2011_Toma_apj_v732.p77..77,Geng_2018_Huang_apjs_v234.p3..3}).

We set the GRB jet as an on-axis-observed top-hat jet without considering the lateral expansion.
The effect of the equal-arrival-time surface (EATS) is considered (e.g., \citealp{Waxman_1997_ApJ...485L...5W}).
We divide the radiation surface as small rings for the on-axis-observed jet.
The intrinsic SEDs can be obtained from the integration over the EATS after considering the Doppler boosting effect.
The EBL absorption effect is taken into account for calculating the observed high energy photons
\citep{Dominguez_2011_Primack_mnras_v410.p2556..2578}.
By summing the flux from each ring observed at a same observer time, the total observed flux can be obtained.

\begin{deluxetable*}{cccccccc}[htbp]
\label{tab1}
\tablecaption{Multiband Fitting Results with the Forward Shock Model}
\tablehead{
\colhead{GRB} & \colhead{${\rm log_{10}}E_{\rm k, iso}~({\rm erg}$)}
&\colhead{${\rm log_{10}}\Gamma_{0}$} &  \colhead{$p$}
& \colhead{${\rm log_{10}}\epsilon_{e}$} & \colhead{${\rm log_{10}}\epsilon_{B}$}
& \colhead{${\rm log_{10}}\theta_{j}$} & \colhead{${\rm log_{10}}A_{\star}$}}
\startdata
221009A & $54.83^{+0.10}_{-0.08}$  & $ 2.28^{+0.02}_{-0.03}$ & $2.60^{+0.03}_{-0.02}$
& $-0.69\pm 0.02$ & $-2.73^{+0.07}_{-0.10}$ & $-1.61^{+0.06}_{-0.06}$ & $-1.91^{+0.01}_{-0.01}$
\enddata
\end{deluxetable*}

\begin{figure*}[htbp]
\centering
\includegraphics[width=0.9\textwidth]{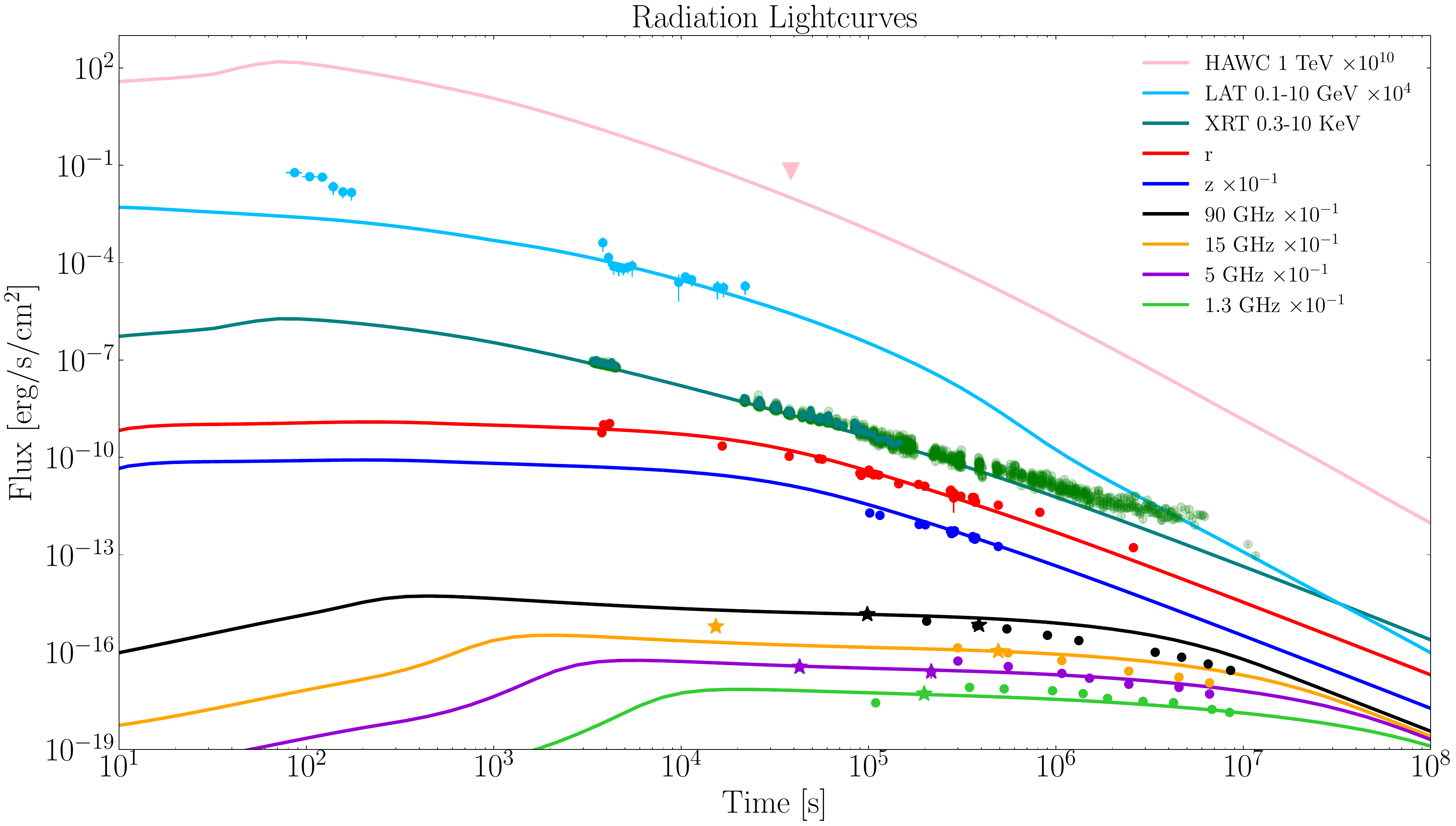}
\caption{
Multiband lightcurves from radio to TeV band plotted with our best fitting result.
The jet generation time is set at $T_0+220$~s with $T_0$ being the trigger time of {\em Fermi}/GBM.
The optical data have performed the Galactic extinction correction, and the
light curve at 1~TeV has considered the EBL absorption.
We show the HAWC 1~TeV upper limit at $T_0+8$~hours.
We have added newly publicly available data released during the review process.
For easy differentiation, the radio data used in our fitting are marked by stars,
and the latest X-ray data are indicated by transparent circles.
Although the optical and X-ray data suggest that model curves need to have a larger jet opening-angle,
the different slopes between the X-ray and radio band data are systematic so
that they cannot be explained by the standard stellar wind environment model.}
\label{fig_LC}
\end{figure*}

\begin{figure*}[htbp]
\centering
\includegraphics[width=0.9\textwidth]{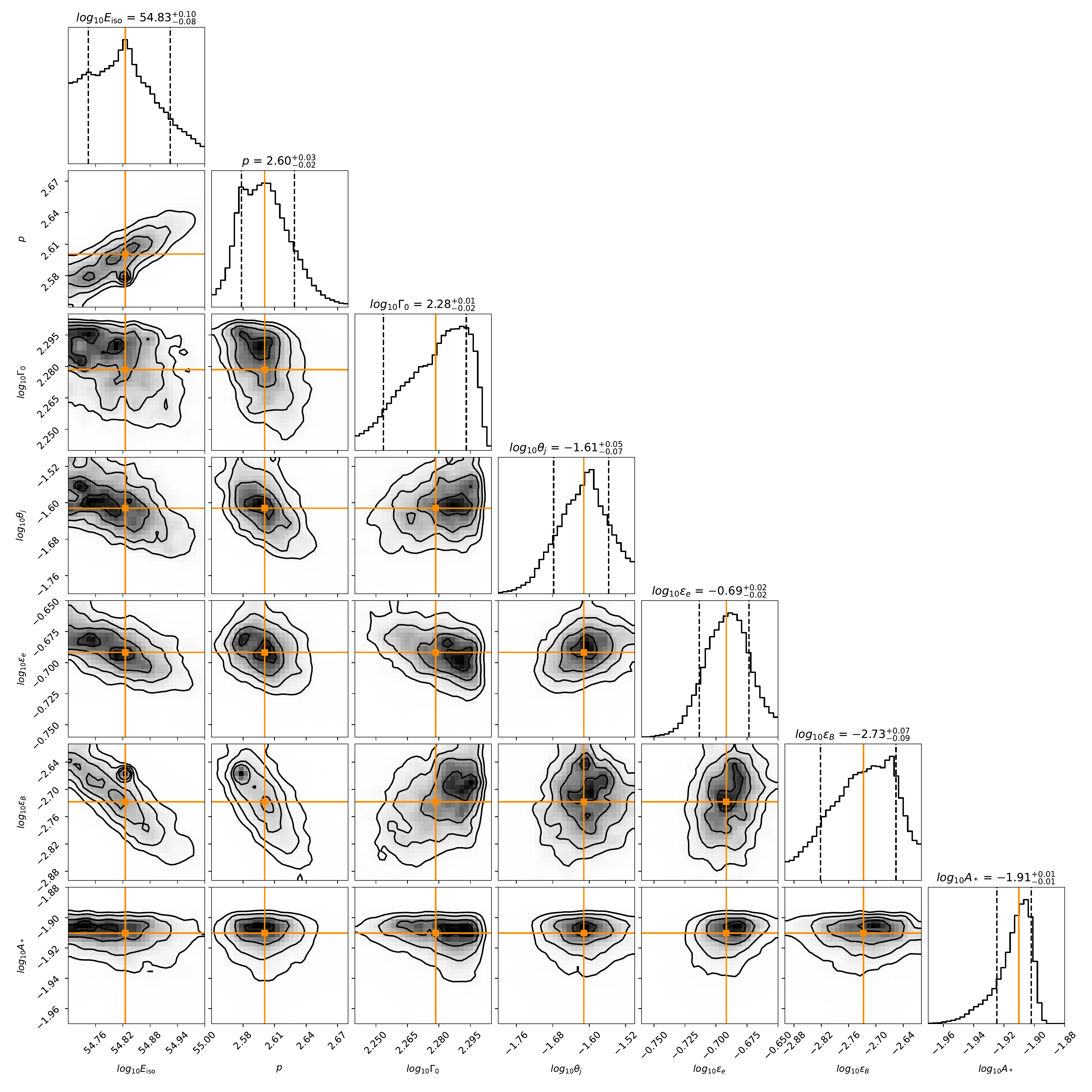}
\caption{Contour plot for the posterior distributions of parameters.}
\label{fig_posterior}
\end{figure*}

\subsection{Inference Results}
We use the {\tt emcee} Python package \citep{ForemanMackey_2013_Hogg_pasp_v125.p306..312}
to fit the lightcurves from radio to GeV bands and the spectra of
{\em Swift}/XRT and {\em Fermi}/LAT (see Section~{\ref{SED}}).
The parameter inference has been performed,
including the isotropic kinetic energy $E_{\rm k, iso}$, the electron energy fraction $\epsilon_{e}$, the
magnetic enerfy fraction $\epsilon_{B}$, $p$, the initial bulk Lorentz factor $\Gamma_0$,
the half-opening angle of jet $\theta_{j}$, and the wind parameter $A_\star$.
Note that the number density of medium $n(r)=3\times 10^{35}A_\star r^{-2}~\rm cm^{-3}$.
Considering that the precursor radiation is observed when the GBM trigger,
the energy of precursor is not sufficient to drive a powerful jet.
Hence we set the generation time of an energetic jet as $\sim T_0+220$~s,
which is the time of the main burst start.
Furthermore, we have also considered the Galactic extinction in the $r$ and $z$ bands
which are $A_{r}=3.52$~mag and $A_{z}=1.95$~mag, respectively
\citep{Schlafly_2011_Finkbeiner_ApJ...737..103S}.
Additionally, extinction from host galaxy $E(B-V)_{\rm host}=0.185$
has chosen with $R_{V,\rm host}=2.93$
(SMC, \citealp{Pei_1992_ApJ...395..130P,Kann_2023_Agayeva_arXiv230206225K}).

We have examined the possibility of GRB~221009A that took place
in a typical interstellar medium (ISM) environment.
We find that for reasonable parameters the model curves of on-axis viewed top-hat jet
are always higher than the observed data reported in radio bands,
say, the radio observations strongly reject the ISM environment under
this scenario.
As a result, we find that the multiband afterglow of GRB~221009A
can be explained by a relativistic jet propagating in
the circumburst environment dominated by a stellar wind \citep{Dai_1998_Lu_mnras_v298.p87..92}.
Our fitting results are shown in Figure~\ref{fig_LC}.
The derived model parameters are reported in Table~\ref{tab1},
and their posterior distributions are shown in Figures~\ref{fig_posterior}.

It shows that the initial $\sim T_0+400$~s GeV emission observed by
{\em Fermi}/LAT in Figure~\ref{fig_LC} may be dominated by the prompt emission.
Only the long lasting emission in the time range from $T_0+3000$~s to $T_0+20000$~s
can be well explained by the external-forward shock radiation from the jet only.
Obviously, our model explains the early time radio-to-GeV band data well.
The model light curves in radio bands are consistent with the reported observations\footnote{
The latest observations in radio bands have been reported during the review of this paper.
Our predicted curves are beyond the data and cannot be corrected by further fitting,
which suggest more considerations. See discussion in Section~\ref{Discussion}.}.
It is worth noting that the excess of the model light curve to $\sim 15$~GHz band radio observations at early times
is possibly due to the radiation from a reverse shock \citep{GCN32676}.
We show the HAWC $95\%$ upper limit on the flux at 1~TeV in Figure~\ref{fig_LC},
and find that our result satisfies this constraint.

\begin{figure}[htbp]
\centering
\includegraphics[width=0.5\textwidth]{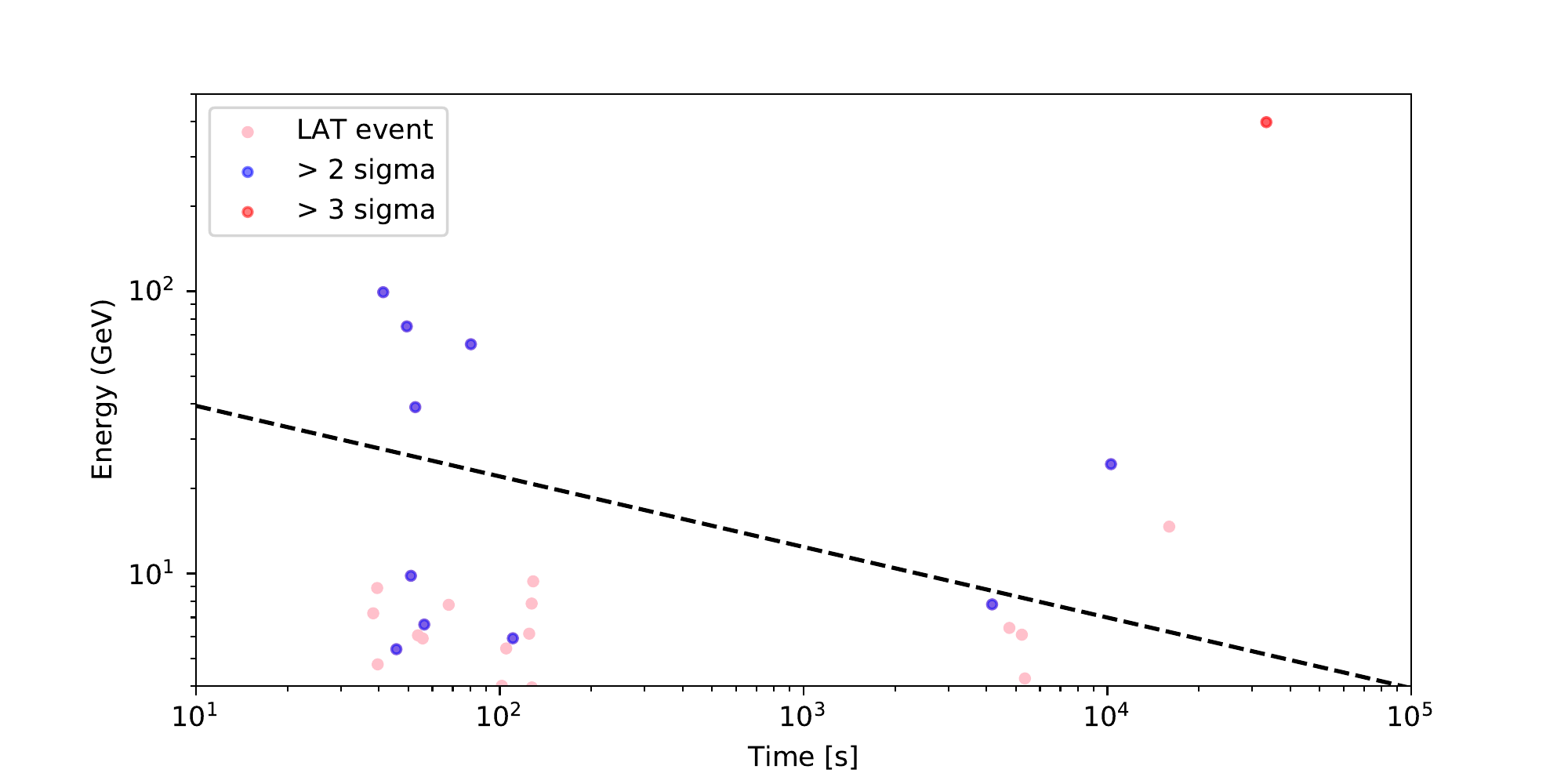}
\caption{The red, blue, and pink dots represent photon events ($>5$~GeV, within $1^{\circ}$) with different probabilities originating from GRB~221009A, respectively.
The photon energy of the only red dot is 397.7~GeV, which is consistent with the result of \cite{GCN32748}.
The black dashed line is the maximum photon energy that synchrotron radiation can reach.}
\label{fig_synmaxe}
\end{figure}

The jet of GRB~221009A towards us  seems to propagate in a wind dominated environment
with no significant density jump in the radial profile.
Based on our inference result, the parameter describing the intensity of the wind $A_{\star}=1.2\times 10^{-2}$,
shows that the progenitor had a relatively small outflow of material prior to the burst.
The microscopic physical parameters $\epsilon_e=0.2$ and $\epsilon_B=1.86\times10^{-3}$
lie within the typical value ranges.
We notice that the half-opening angle of jet
$\theta_{j}\simeq1^{\circ}.4$ is mainly determined by
$z$ band data we used, which
may be changed by observational data with better qualities.
The combination of parameters is not exceptional,
suggesting that GRB~221009A could be a typical wind environment dominated long GRB.
We calculate the efficiency of prompt emission to the total enegy of jet as
$\eta_\gamma =E_{\gamma,\rm iso}/(E_{\gamma,\rm iso}+E_{\rm k, iso})\sim 69\%$ \citep{An_2023_arXiv230301203A,Yang_2023_Zhao_arXiv230300898Y}
which is consistent with the expectation of the fireball model
(e.g., \citealp{Lloyd_Ronning_2004_Zhang_ApJ...613..477L, Zhang_2007_Liang_ApJ...655..989Z}).
Furthermore, we find that the inclusion of GeV data partly
breaks the parameter degeneracy commonly observed in GRB afterglow fitting \citep[e.g.,][]{Ren_2020_Lin_apj_v901.p26..26L},
underlining the significance of high-energy data in determining accurate parameters for GRB afterglows.
Based on the derived model parameters, we also calculate the maximum photon energy generated by the synchrotron radiation \citep{cheng1996spectral,fan2013high}, which indicates that the high-energy photons come from inverse Compton radiation by comparing the {\em Fermi}/LAT observations (see Figure~\ref{fig_synmaxe}).

\subsection{Spectral Energy Distribution}\label{SED}
\begin{figure*}[htbp]
\centering
\includegraphics[width=0.95\textwidth]{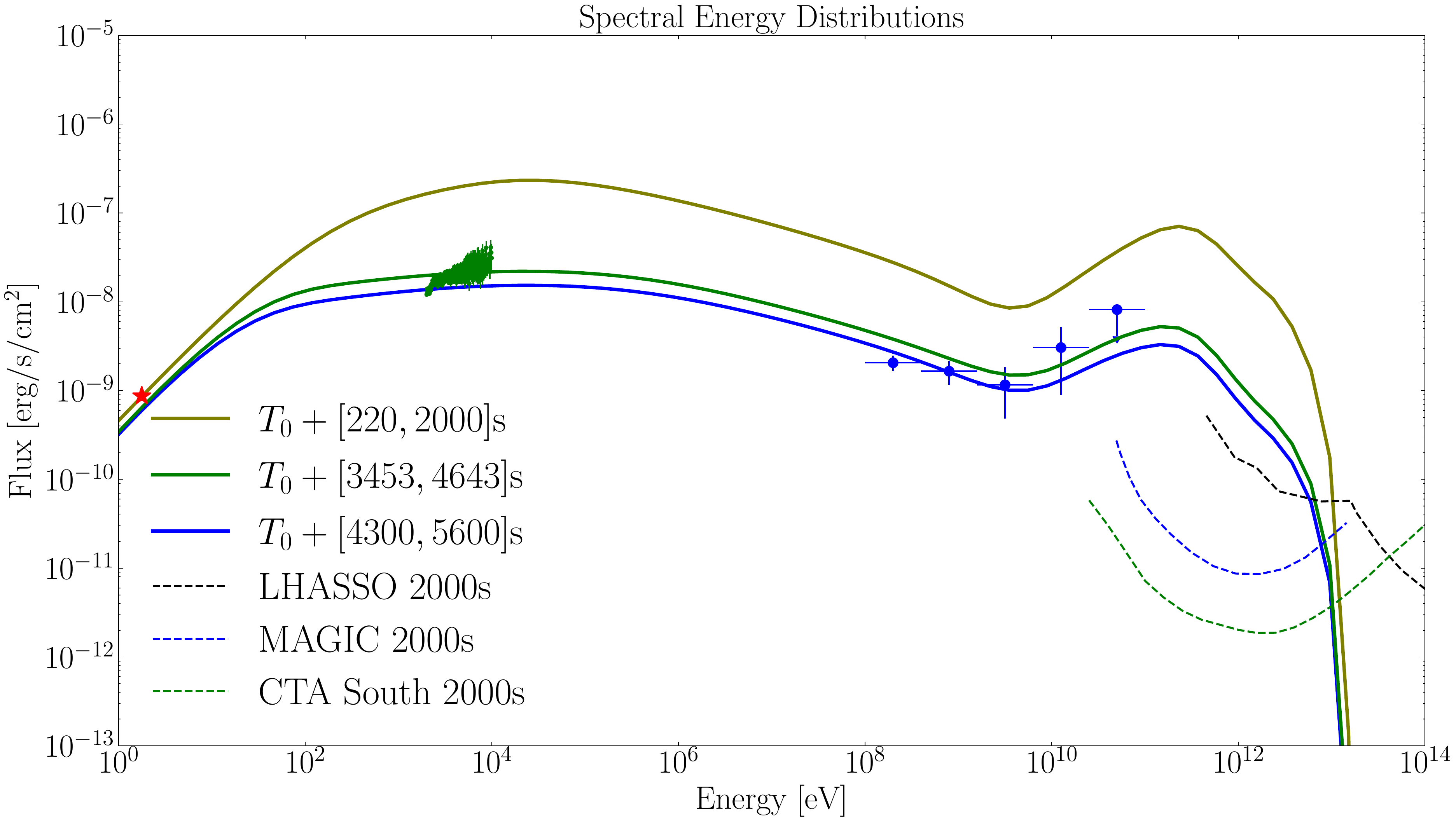}
\caption{Afterglow SEDs of three chosen timeslices.
The red star marks the $R$ band data at $T_0+4150$~seconds from \cite{GCN32645},
the green points mark the $2-10$~keV {\em Swift}/XRT data within $T_0+[3453,4643]$~s, and
the blue points mark the $0.1-10$~GeV {\em Fermi}/LAT data within $T_0+[4300,5600]$~s, respectively.
The black, blue, and green dashed lines correspond to the detection sensitivities of LHASSO, MAGIC, and CTA in 2000~seconds, respectively \citep{Cao_2019_dellaVolpe_arXiv190502773C}.}
\label{fig_spec}
\end{figure*}

By using the parameters of fitting results,
we show the SEDs of GRB~221009A afterglow in different timeslices,
see Figure~\ref{fig_spec}.
We show the observed spectral data and corresponding fitting curves
within $T_0+[4300,5600]$~s for {\em Fermi}/LAT,
and $T_0+[3453,4643]$~s for {\em Swift}/XRT, respectively.
We also plot the SED of $T_0+[220,2000]$~s as a prediction result for the LHASSO observation spectrum.
One can observe that the fit between the theoretical line
and the observations is quite good in the X-ray and GeV bands.
Comparing the sensitivity curves of LHASSO, MAGIC, and CTA
with the SED in our model, we find that the radiation in the $0.1-10$~TeV band
is extremely bright for detections during $T_0+[220,2000]$~s.
Our model gives the maximum SSC radiation flux around 300~GeV,
with the value of $\sim 10^{-7}~\rm erg~cm^{-2}~s^{-1}$.
We find that the VHE radiation of GRB 221009A has been cut off
around 10~TeV as a result of EBL absorption up to the redshift $z=0.151$ \citep{Dominguez_2011_Primack_mnras_v410.p2556..2578}.
Compared with the sensitivity curve of LHASSO,
the highest energy photons from the SSC process with energy of
11~TeV can be detected by LHASSO with 2000~seconds exposure
by using the EBL model of \cite{Dominguez_2011_Primack_mnras_v410.p2556..2578}.
If the 18~TeV photon detected by LHASSO at $T_0+2000$~s was realistic,
it is hard to explain by the SSC mechanism only
and suggest the existence of other radiation processes.
Considering that EBL absorption has a rate of
$e^{-\tau_{\rm EBL}}\sim 10^{-8}$ at $18$~TeV,
the intrinsic flux required to account for the radiation process of this photon event
is approximately larger than $10^{-2}~\rm erg~cm^{-2}~s^{-1}$
for the estimated sensitivity thresholds of LHASSO.
This is corresponding to the radiation luminosity greater than
$\sim 6.5\times 10^{53}~\rm erg~s^{-1}$ at $\sim 18$~TeV.
Some possibilities can explain this phenomenon,
such as hadronic processes (e.g., synchrotron radiation of protons, \citealp{Aharonian_2000_NewA....5..377A,Alves_2018_Zrake_PhRvL.121x5101A}),
Lorentz invariance violation (LIV) and Axion-like particles \citep[e.g.,][]{Galanti_2022_Roncadelli_arXiv221106935G,Nakagawa_2022_Takahashi_arXiv221010022N,
Finke_2023_Razzaque_ApJ...942L..21F,Li_2023_Ma_AstroparticlePhysics_v148.p102831..102831},
or the need for corrections to the EBL field of low-energy photons.

\section{Discussion}\label{Discussion}
During the review process of this paper, several pieces of data were public.
The radio afterglow of GRB~221009A exhibits a strange behavior on its spectrum,
which is difficult to explain from a standard ISM or wind environment perspectively
\citep[see][for detailed discussion]{Laskar_2023_Alexander_arXiv230204388L}.
Figure~\ref{fig_LC} contains the radio data from
\cite{Laskar_2023_Alexander_arXiv230204388L}
confirming that our model is inconsistent with the radio afterglow.
On the other hand,
the spectroscopy from JWST, HST and X-shooter shows flat NIR spectra
\citep[][where they proposed $p<2$]
{Levan_2023_Lamb_arXiv230207761L,Malesani_2023_Levan_arXiv230207891M},
again implying a violation of the standard model.
Regardless, the data and analysis released so far seem to indicate that GRB~221009A
exhibits a more complex radiation behavior than was expected
based on a limited amount of information available.

The broadband afterglows of GRB~221009A
have been explained in several papers using structured jet models
\citep[e.g.,][]{Sato_2022_Murase_arXiv221209266S,O'Connor_2023_Troja_arXiv230207906O}.
The structured jet picture has been proposed \citep[e.g.,][]{Meszaros_1998_Rees_ApJ...499..301M,Dai_2001_Gou_ApJ...552...72D,
Zhang_2002_Meszaros_ApJ...571..876Z,Rossi_2002_Lazzati_MNRAS.332..945R}
and supported by GRB~170817A \citep[e.g.,][]
{Troja_2017_Piro_Natur.551...71T,Lazzati_2018_Perna_PhRvL.120x1103L,
Ren_2020_Lin_apj_v901.p26..26L,Makhathini_2021_Mooley_ApJ...922..154M,
Balasubramanian_2022_Corsi_ApJ...938...12B}.
Complex geometry produces the atypical spectral behaviors,
but it also introduces a number of free parameters.
Recent works invoking the structured jet to interpret GRB~221009A
have not yet added a full fit to the multiband afterglows,
which may undermine the credibility of the model.
Other possible explanations imply some changes in microscopic parameters
(e.g., $p$, $\epsilon_e$, $\epsilon_B$)
corresponding to the shock acceleration behavior and the dissipative mechanism of
energy within the shocked material that is still poorly understood
\citep[e.g.,][]{Laskar_2023_Alexander_arXiv230204388L,Levan_2023_Lamb_arXiv230207761L,
Malesani_2023_Levan_arXiv230207891M}.

It is possible to distinguish the correct direction of the model improvement
based on the early afterglow data and future observations.
Due to the high zenith Angle at the time of the burst,
LHASSO was able to record the high-energy afterglow of GRB~221009A
in a highly sensitive state.
This may help us to distinguish between different models.
As has been pointed out, the rising slope of the early afterglow
would give a strong constraint on the wind model.
A typical stellar wind environment gives
the early light curve that is different from the behavior predicted in an ISM environment
\citep[e.g.,][]{Wu_2004_Dai_ChJAA...4..455W,Wu_2005_Dai_ApJ...619..968W,
Fan_2008_Piran_mnras_v384.p1483..1501,Zhao_2022_Cheng_Univ....8..588Z}.
However, considering the long-duration prompt emission of GRB~221009A,
we want to point out the possible influence of
prompt radiation on the early-time observations,
for instance, the EIC processes and the internal hadronic dissipation
\citep{Zhang_2022_Murase_arXiv221105754Z,
Rudolph_2023_Petropoulou_ApJ...944L..34R,Wang_2023_Ma_arXiv230211111W},
and even the early energy injection \citep{Li_2023_Lin_ApJ...944...21L}.
In addition, possibly existing reverse shock may also play an important role
\citep{Kobayashi_2003_Zhang_ApJ...597..455K,Wu_2003_Dai_MNRAS.342.1131W},
but there has been no early afterglow data released yet.
No matter whether a complex jet structure or a modification of the radiation process is preferred,
studies on GRB~221009A will provide us with deeper insights.

\section{Summary \& Conclusions}\label{Summary}
In this paper, we have modeled the dynamics and radiation physics of the afterglow
of the brightest GRB~221009A in detail.
The afterglow of GRB~221009A can be explained by a relativistic jet
with an initial Lorentz factor $\Gamma_0\simeq190$ propagating in an environment
dominated by a stellar wind with parameter $A_\star =1.2\times 10^{-2}$.
Due to the lack of data of early onset bump, these are reference values as current estimates.
We explained the afterglow lightcurves and SEDs of GRB~221009A
by using the synchrotron radiation plus the SSC mechanism.
Although we have well fitted the first week radio-to-GeV band observations,
the late time radio data is hardly explained under the standard external forward shock model
of a top-hat jet in a stellar wind environment.
According to \cite{Laskar_2023_Alexander_arXiv230204388L},
a modification to the mechanism of relativistic synchrotron radiation
or an additional population of electrons will be necessary
to interpret radio data under the standard model.
The other explanations like structured jet models with complex geometry are also hopeful \citep[e.g.,][]{Sato_2022_Murase_arXiv221209266S,O'Connor_2023_Troja_arXiv230207906O}.

In this paper, we have presented the prediction of SSC composition
under the standard stellar wind environment model as a useful reference.
Our results show that the SSC component with photon energy large than $100$~GeV
is extremely bright during the first $\sim 2000$~seconds
after the {\em Fermi}/GBM trigger time $T_0$ of GRB~221009A.
The time integral SED within $T_0+220$~s to $T_0+2000$~s
shows the SSC component has peaked in $\sim 300$~GeV
and the peak flux $\nu F_{\nu}(300~\rm GeV) \sim 10^{-7}~\rm erg~cm^{-2}~s^{-1}$.
Spectra above 300~GeV corrected with the EBL
show an approximate power-law distribution up to 1~TeV,
and rapidly drops with photon energies greater than 1~TeV
before a cut off at approximately 10~TeV.
Our results indicate that the SSC component is very promising
for the interpretation of VHE emission from GRB~221009A.
If the $18$~TeV event detected by LHASSO is realistic,
we suggest an additional component (relative to the SSC component)
to account for photons beyond $10$~TeV,
e.g., synchrotron radiation of protons.
The corresponding luminosity of such a radiation process
is required to have at least
$\nu L_{\nu}\sim 6.5\times 10^{53}~\rm erg~s^{-1}$ at $18$~TeV.
Also, other possible reasons such as Lorentz invariance violation (LIV)
or corrections to the EBL field needs to be examined.
Furthermore, we find that the inclusion of GeV data partly breaks
the parameter degeneracy commonly observed in GRB afterglow fitting
(e.g., \citealp{Ren_2020_Lin_apj_v901.p26..26L}),
underlining the significance of high-energy data
in determining accurate parameters for GRB afterglows.

\acknowledgments
We thank the anonymous referee for the helpful comments and suggestions to improve this work.
We thank D. A. Kann and T. Laskar for useful comments.
This work was supported by
the National Natural Science Foundation of China (grant No. 11833003).
This work used data and software provided by the Fermi Science Support Center and data supplied by the UK Swift Science Data Centre at the University of Leicester. This research also used of the CTA instrument response functions provided by
the CTA Consortium and Observatory, see \url{http://www.cta-observatory.org/science/cta-performance/} (version prod3b-v2) for more details.


\software{\texttt{Matplotlib}
\citep{Hunter_2007__ComputinginScienceandEngineering_v9.p90..95},
\texttt{Numpy}
\citep{Harris_2020_Millman_Nature_v585.p357..362},
\texttt{emcee}
\citep{ForemanMackey_2013_Hogg_pasp_v125.p306..312},
\texttt{corner}
\citep{ForemanMackey_2016__TheJournalofOpenSourceSoftware_v1.p24..24},
\texttt{Astropy}
\citep{Collaboration_2013_Robitaille_aap_v558.p33..33A}}
\clearpage


\end{document}